\documentclass[aps, twocolumn, floatfix, nofootinbib, showpacs, reprint, 10pt]{revtex4-1}
\usepackage[T1]{fontenc}
\usepackage{amsmath}
\usepackage{txfonts}
\usepackage{graphicx}
\usepackage{xspace}
\usepackage[colorlinks, urlcolor=black, citecolor=black, link color=black]{hyperref}
\newcommand{\bracket}[3]{\langle #1 \lvert #2 \rvert #3 \rangle}

\newcommand{\modulus}[1]{\left| #1 \right|}
\def \Re{\text{Re}}
\def \Im{\text{Im}}

\def\Dbar {\kern 0.2em\overline{\kern -0.2em D}{}\xspace}

\def\Dzb   {\ensuremath{\Dbar^0}\xspace}
\def\DzDzb {\ensuremath{D^0 {\kern -0.16em - \Dzb}}\xspace}
\def\Dp    {\ensuremath{D^+}\xspace}

\def\Dm    {\ensuremath{D^-}\xspace}
\def\Dz    {\ensuremath{D^0}\xspace}

\def\Dzbs   {\ensuremath{\Dbar^{*0}}\xspace}
\def\DzsDzbs {\ensuremath{D^{*0} {\kern -0.16em \Dzbs}}\xspace}

\def\Bbar  {\kern 0.18em\overline{\kern -0.18em B}{}\xspace}

\def\Bzb   {\ensuremath{\Bbar^0}\xspace}
\def\BzBzb {\ensuremath{B^0 {\kern -0.16em -\Bzb}}\xspace}
\def\Bp    {\ensuremath{B^+}\xspace}

\def\Bm    {\ensuremath{B^-}\xspace}
\def\Bz    {\ensuremath{B^0}\xspace}
\def\Bbar  {\kern 0.18em\overline{\kern -0.18em B}{}\xspace}

\def\Bzb   {\ensuremath{\Bbar^0}\xspace}
\def\BzBzb {\ensuremath{B^0 {\kern -0.16em -\Bzb}}\xspace}
\def\ubar	{\ensuremath{\overline{u}}\xspace}

\def\cbar	{\ensuremath{\overline{c}}\xspace}

\def\bbar	{\ensuremath{\overline{b}}\xspace}

\begin{document}

\title{Probing new physics in \texorpdfstring{$B \to D \ell^+ \ell^-$}{B -> D l+
l-} decays by using angular asymmetries} %

\author{C.~S.~Kim} \email[E-mail at: ]{cskim@yonsei.ac.kr}%
\affiliation{Department of Physics and IPAP, Yonsei University, Seoul 120-749,
Korea} %

\author{Dibyakrupa Sahoo} \email[E-mail at: ]{sahoodibya@yonsei.ac.kr}%
\affiliation{Department of Physics and IPAP, Yonsei University, Seoul 120-749,
Korea} %

\date{\today}

\begin{abstract}
We present the fully general, model independent study of a few rare semileptonic
$B$ decays that get dominant contributions from $W$-annihilation and
$W$-exchange diagrams, in particular $\Bz \to \Dzb \ell^+\ell^-$, where $\ell =
e,\mu$. We consider the most general Lagrangian for the decay, and define three
angular asymmetries in the Gottfried-Jackson frame, which are sensitive to new
physics. We show how these angular asymmetries can be easily extracted from the
distribution of events in the Dalitz plot for $B \to D \ell^+ \ell^-$ decays.
Especially a non-zero forward-backward asymmetry within the frame would give the
very first hint of possible new physics. These observations are also true for
related decay modes, such as $\Bp \to \Dp \ell^+ \ell^-$ and $\Bz \to \Dz \ell^+
\ell^-$. Moreover, these asymmetry signatures are not affected by either
$\BzBzb$ or $\DzDzb$ mixings. Then, this implies that both $\Bz \to \Dzb \ell^+
\ell^-$ and $\Bz \to \Dz \ell^+ \ell^-$ as well as their CP conjugate modes can
all be considered together in our search for signature of new physics. Hence, it
would be of great importance to look for and study these decays in the
laboratory, LHCb and Belle II in particular.
\end{abstract}


\maketitle

\section{Introduction}\label{sec:intro}

It is very well known that despite having enormous success in explaining an
astounding amount of experimental observations, the Standard Model (SM) of
particle physics has many glaring lacunae. On the other hand, with little direct
experimental validation, there exists a vast possibility of physics beyond the
SM. In addition to the direct collider searches, experimenters and theorists
alike have tried to look for physics beyond the SM in various meson decays, most
notably in the $B \to K^* \ell^+ \ell^-$ decays~\cite{Aaij:2015oid,
Karan:2016wvu, Mandal:2015bsa, Mandal:2014kma, Altmannshofer:2014rta,
Ciuchini:2015qxb, Jager:2014rwa, Descotes-Genon:2015uva, Sahoo:2015qha}, where
$\ell = e, \mu$. In such exciting times for new physics (NP), we present the
most general, model independent study of a few hitherto
unseen~\cite{Agashe:2014kda}, very rare, semileptonic decays of the $B$ meson,
namely $\Bz \to \Dzb \ell^+ \ell^-$, $\Bp \to \Dp\ell^+ \ell^-$ and $\Bz \to \Dz
\ell^+ \ell^-$, which are promising candidates to probe new physics. WE note
that some of the previous works, which dealt with decay modes similar to those
we consider now, are given in \cite{Evans:1999kv, Evans:1999wx, Lebed:1999kq,
Evans:1999zc}. Unlike previous works, we provide a completely model-independent
study of $B \to D\ell^+ \ell^-$ decays by using the effective field theory
framework. These decay modes are primarily facilitated by $W$-annihilation and
$W$-exchange diagrams and are hence, in general, highly suppressed. However, for
large Wilson coefficients such decays (especially $\Bz \to \Dzb \ell^+ \ell^-$)
can have sizeable branching ratios in the SM, $\mathcal{O}\left(10^{-5}\right)$
\cite{Kim:2011ps}, such that they can be observed and studied experimentally.
The decays $\Bz \to \Dz \ell^+ \ell^-$ and $\Bp \to \Dp \ell^+ \ell^-$ being CKM
suppressed\footnote{In $\Bz \to \Dzb \ell^+ \ell^-$ the two quark transitions
are $\bbar \to \cbar W^+$ (which is CKM-suppressed) and $d \to u W^-$ (which is
CKM-favoured). However, in $\Bp \to \Dp \ell^+ \ell^-$ and $\Bz \to \Dz \ell^+
\ell^-$ the two quark transitions are $\bbar \to \ubar W^+$ and $d \to c W^-$,
both of which are CKM-suppressed.} have smaller branching ratios in the SM,
$\mathcal{O} \left(10^{-9}-10^{-11}\right)$, which has been estimated without
considering the photon pole contribution \cite{Evans:1999kv,Evans:1999zc}.
However, assuming that these rare decays can be observed experimentally in near
future, we provide angular observables which are easy to study in experiments
and show how they can be utilised to unearth any underlying new physics in these
decay modes. It is also notable that the signatures of new physics, considered
later in this work, are unaffected by $\BzBzb$ or $\DzDzb$ mixings. This
therefore allows us to combine the data sets for the decay modes $\Bz \to \Dzb
\ell^+ \ell^-$, $\Bz \to \Dz \ell^+ \ell^-$ and their CP conjugate modes and do
a search for the signatures of new physics with more statistics. We are hopeful
that our results would further motivate the experimentalists to look for these
decays in the various ongoing and upcoming particle physics experiments, such as
LHCb and Belle II.

This paper is organised as follows: In section~\ref{sec:lagrangian-amplitude} we
provide the model-independent form of the most general Lagrangian which then
dictates the form of the decay amplitude. This is followed by a discussion on
the relevant kinematics in the Gottfried-Jackson frame. Using the most general
amplitude, we find out the angular distribution in
section~\ref{sec:general-angular-distribution}. We also provide the necessary
angular asymmetries which are sensitive to specific parts of the angular
distribution. Finally, we analyse the angular asymmetries in detail and show in
section~\ref{sec:new_physics_signatures} how they can be used to decipher the
signature of any new physics contributing to the processes under consideration.
In section~\ref{sec:expt-sign} we show how these asymmetries can be very easily
obtained from the distribution of events inside the Dalitz plots for $B \to D
\ell^+ \ell^-$ decays. In section~\ref{sec:mixing-independence} we note that the
signatures of new physics are independent of $\Bz$-$\Bzb$ and $\Dz$-$\Dzb$
mixing. Then we conclude in section~\ref{sec:conclusion} emphasizing all the
essential aspects of our analysis.

\section{The most general Lagrangian and Amplitude}\label{sec:lagrangian-amplitude}

The model-independent effective Lagrangian contributing to the $B \to D \ell^+
\ell^-$ decays can be written as follows,
\begin{align}
\mathcal{L}_{\textrm{eff}} &= J_S \left( \overline{\ell}\mathbf{1} \ell \right)
+ J_P \left( \overline{\ell} \gamma^5\ell \right) + \left(J_V\right)_{\alpha}
\left( \overline{\ell}\gamma^{\alpha}\ell \right) + \left(J_A\right)_{\alpha}
\left( \overline{\ell}\gamma^{\alpha}\gamma^5\ell \right) \nonumber\\%
&\quad + \left(J_{T_1}\right)_{\alpha\beta} \left(
\overline{\ell}\sigma^{\alpha\beta}\ell \right) +
\left(J_{T_2}\right)_{\alpha\beta} \left(
\overline{\ell}\sigma^{\alpha\beta}\gamma^5 \ell \right),
\label{eq:Effective-Lagrangian}
\end{align}
where $J_S$, $J_P$, $\left(J_V\right)_{\alpha}$, $\left(J_A\right)_{\alpha}$,
$\left(J_{T_1}\right)_{\alpha\beta}$, $\left(J_{T_2}\right)_{\alpha\beta}$ are
the different effective hadronic currents which effectively describe the
quark-level transitions from $B$ to $D$ meson. In the SM, only the vector and
axial vector currents contribute. So all other terms in
Eq.~\eqref{eq:Effective-Lagrangian} except $\left(J_V\right)_{\alpha}$ and
$\left(J_A\right)_{\alpha}$ are possible only in case of some specific NP
scenarios. In this work, we are not concerned about which particular kind of NP
model would give rise to such terms, though such model-dependent approaches are
also fruitful. While specific NP models would give specific signatures of NP, we
are interested in finding out the generic signature of NP in this work, starting
with Eq.~\eqref{eq:Effective-Lagrangian}. It must be noted that NP can also
modify both $\left(J_V\right)_{\alpha}$ and $\left(J_A\right)_{\alpha}$ from
their SM expressions.

In order to get the most general amplitude, we need to go from the effective
quark-level description of Eq.~\eqref{eq:Effective-Lagrangian} to the meson
level description by defining appropriate form factors. It is easy to write down
the most general form of the amplitude for $B \to D \ell^+ \ell^-$ as shown
below,
\begin{align}
\mathcal{M} \left( B \to D \ell^+ \ell^- \right) &= F_S \left(
\overline{\ell}~\mathbf{1}~\ell \right) + F_P \left(
\overline{\ell}~\gamma^5~\ell \right) \nonumber\\*%
&\quad + \left( F_V^+ p_{\alpha} + F_V^- q_{\alpha} \right) \left(
\overline{\ell}~\gamma^{\alpha}~\ell \right) \nonumber\\* %
&\quad + \left( F_A^+ p_{\alpha} + F_A^- q_{\alpha} \right) \left(
\overline{\ell}~\gamma^{\alpha}~\gamma^5~\ell \right) \nonumber\\* %
&\quad + F_{T_1}~p_{\alpha}~q_{\beta} \left(
\overline{\ell}~\sigma^{\alpha\beta}~\ell \right) \nonumber\\* %
&\quad + F_{T_2}~p_{\alpha}~q_{\beta} \left(
\overline{\ell}~\sigma^{\alpha\beta}~\gamma^5~\ell \right), \label{eq:amplitude}
\end{align}
where $F_{S}$, $F_{P}$, $F_{V}^{\pm}$, $F_{A}^{\pm}$, $F_{T_1}$ and $F_{T_2}$
are the relevant form factors, and are defined as follows,
\begin{subequations}
\begin{align}
\bracket{D}{J_S}{B} &= F_S,\\%
\bracket{D}{J_P}{B} &= F_P,\\%
\bracket{D}{\left(J_V\right)_{\alpha}}{B} &= F_V^+ p_{\alpha} + F_V^-
q_{\alpha},\\%
\bracket{D}{\left(J_A\right)_{\alpha}}{B} &= F_A^+ p_{\alpha} + F_A^-
q_{\alpha},\\%
\bracket{D}{\left(J_{T_1}\right)_{\alpha\beta}}{B} &=
F_{T_1}~p_{\alpha}~q_{\beta},\\%
\bracket{D}{\left(J_{T_2}\right)_{\alpha\beta}}{B} &=
F_{T_2}~p_{\alpha}~q_{\beta},
\end{align}
\end{subequations}
with
\begin{equation}
p \equiv p_B + p_D,\quad q \equiv p_B - p_D,
\end{equation}
in which $p_B$ and $p_D$ are the 4-momenta of the $B$ meson and $D$ meson
respectively. All the form factors appearing in the amplitude are complex, in
general, and contain all information regarding any new physics. Terms containing
$F_V$ and $F_A$ are the ones allowed in the SM. As noted before NP can also
alter $F_V$ and $F_A$ in addition to introducing other form factors. We shall
use the angular distribution of $B \to D \ell^+ \ell^-$ decays to find out the
various signatures of NP.

\begin{figure}[hbtp]
\centering %
\includegraphics[scale=0.8]{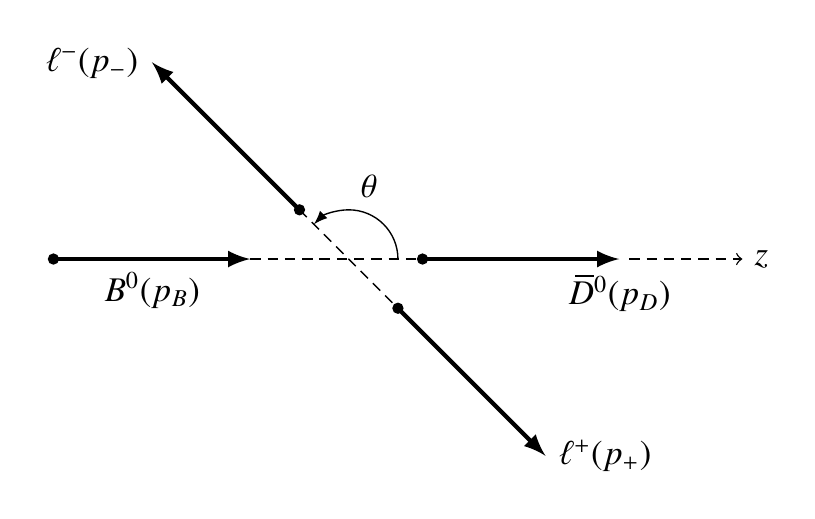} %
\caption{\label{fig:GJ-frame}Decay of $\Bz \to \Dzb \ell^+ \ell^-$ in the
Gottfried-Jackson frame.} %
\end{figure}

We shall discuss the decay $B \to D \ell^+ \ell^-$ in the Gottfried-Jackson
frame, which is shown in Fig.~\ref{fig:GJ-frame}. In this frame the $B$ meson
flies along the positive $z$-direction with 4-momentum $p_B = \left(E_B,
\mathbf{p}_B \right)$ and decays to a $D$ meson which flies along the positive
$z$-direction with 4-momentum $p_D = \left( E_D, \mathbf{p}_D \right)$ and to
$\ell^+$, $\ell^-$ which fly back-to-back with 4-momenta $p_+ = \left( E_+,
\mathbf{p}_+ \right)$ and $p_- = \left( E_-, \mathbf{p}_- \right)$ respectively, such
that by conservation of 4-momentum we get, $\mathbf{p}_+ + \mathbf{p}_- = \mathbf{0}$,
$\mathbf{p}_B = \mathbf{p}_D$, and $E_B = E_D + E_+ + E_-$. The $\ell^-$ flies
outwards subtending an angle $\theta$ with respect to the direction of flight of
the $B$ meson, in the Gottfried-Jackson frame. Let us also denote the three
invariant mass-squares as follows,
\begin{subequations}
\begin{align}
s &= (p_+ + p_-)^2 = (p_B - p_D)^2, \\%
t &= (p_D + p_-)^2 = (p_B - p_+)^2, \\%
u &= (p_D + p_+)^2 = (p_B - p_-)^2.
\end{align}
\end{subequations}
It is easy to show that $s + t + u = m_B^2 + m_D^2 + 2 m_{\ell}^2$, where $m_i$
denotes the mass of particle $i$. In the Gottfried-Jackson frame, the
expressions for $t$ and $u$ are given by
\begin{subequations}
\begin{align}
t &= a - b \cos\theta,\label{eq:t}\\%
u &= a + b \cos\theta,\label{eq:u}
\end{align}
\end{subequations}
where
\begin{subequations}\label{eq:ab}
\begin{align}
a &= \left( m_B^2 + m_D^2 + 2 m_{\ell}^2 - s \right)/2,\\ %
b &= \sqrt{\lambda\left( m_B^2, m_D^2, s \right) \left(1 - 4 m_{\ell}^2/s
\right)}~/2,
\end{align}
\end{subequations}
with the K\"{a}ll\'{e}n function $\lambda(x,y,z)$ defined as
\begin{equation}
\lambda\left( x,y,z \right) = x^2 + y^2 + z^2 - 2 \left( xy + yz + zx \right).
\end{equation}
It is clear that both $a$ and $b$ are functions of $s$ only. We would also like
to emphasize that the angular distribution given later in
section~\ref{sec:observable-NP} involves the angle $\theta$ which is the angle
between the directions of flight of $\ell^-$ and $D$ measured in the
Gottfried-Jackson frame, as shown in Fig.~\ref{fig:GJ-frame}.

\section{The model-independent Observables and Signatures of new Physics}\label{sec:observable-NP}

\subsection{Angular distribution and various  observables}\label{sec:general-angular-distribution}

Using the most general form of the amplitude as given in
Eq.~\eqref{eq:amplitude} we can write down the following expression for the most
general angular distribution for the $B \to D \ell^+ \ell^-$ decays.
\begin{equation}
\frac{d^2 \Gamma}{ds \; d\cos\theta} = \frac{b~\sqrt{s} \left( T_0 +
T_1~\cos\theta + T_2~\cos^2\theta \right)}{128 ~ \pi^3 ~ m_B^2 \left( m_B^2 -
m_D^2 + s \right)},\label{eq:diff-decay-rate}
\end{equation}
where we note again that the angle $\theta$ is measured in the Gottfried-Jackson
frame as shown in Fig.~\ref{fig:GJ-frame}, and
\begin{subequations}
\begin{align}
T_0 &= 8 a^2 \Big[ \modulus{F_{A}^+}^2+4 m_{\ell} \left(\modulus{F_{T_1}}^2
m_{\ell}-\Im\left(F_{T_1} F_{V}^+\right) \right)+\modulus{F_{V}^+}^2 \Big]
\nonumber\\ %
&\quad - 16 a \bigg[ \modulus{F_{A}^+}^2 \left(m_D^2-m_{\ell}^2\right)
\nonumber\\%
&\quad - m_{\ell} \bigg( 4 \left(m_D^2+m_{\ell}^2\right) \left(\Im\left(F_{T_1}
F_{V}^{+*}\right) - \modulus{F_{T_1}}^2 m_{\ell}\right) \nonumber\\%
&\quad + 2 m_{\ell} \Re\left( F_A^+ F_A^{-*} \right)+\Re\left( F_A^+ F_P^*
\right) \bigg) + \modulus{F_{V}^+}^2 \left(m_D^2+m_{\ell}^2\right) \bigg]
\nonumber\\ %
&\quad + 2 s \bigg[4 \bigg(m_{\ell} \bigg(m_{\ell} \left( \modulus{F_{A}^-}^2+2
\Re\left( F_A^+ F_A^{-*} \right) \right) \nonumber\\%
&\quad + 4 m_D^2 \left( \Im\left(F_{T_1} F_{V}^+\right)-\modulus{F_{T_1}}^2
m_{\ell} \right) + \Re\left( F_A^- F_P^* \right) \nonumber\\ %
&\quad +\Re\left( F_A^+ F_P^* \right)\bigg)-\modulus{F_{V}^+}^2
m_D^2\bigg)+\modulus{F_{P}}^2+\modulus{F_{S}}^2\bigg] \nonumber\\ %
&\quad + 8 \modulus{F_{A}^+}^2 \left(m_D^2-m_{\ell}^2\right) \left(m_D^2+3
m_{\ell}^2-s\right) \nonumber\\%
&\quad + 8 \bigg[ m_{\ell} \bigg(-\modulus{F_{S}}^2 m_{\ell} \nonumber\\%
& \quad - 2 \left(m_D^2+m_{\ell}^2\right) \bigg(2 \left(m_D^2 +
m_{\ell}^2\right) \left(\Im\left(F_{T_1} F_{V}^+\right)-\modulus{F_{T_1}}^2
m_{\ell} \right) \nonumber\\ %
&\quad + 2 m_{\ell} \Re\left( F_A^+ F_A^{-*} \right) + \Re\left( F_A^+ F_P^*
\right)\bigg)\bigg) \nonumber\\%
&\quad + \modulus{F_{V}^+}^2 \left(m_D^2+m_{\ell}^2\right)^2\bigg],\\ %
T_1 &= 8 b \bigg( 2 m_{\ell} \big( \Im\left(F_{T_2} F_{A}^{+*}\right) \left( 2 a
- 2 \left( m_D^2+m_{\ell}^2 \right) + s \right) \nonumber\\%
&\quad + \Im\left( F_{T_2} F_A^{-*} \right) s + \Re\left( F_V^+ F_S^* \right)
\big) \nonumber\\%
&\quad + s \Big( \Im\left( F_{T_2} F_P^{*} \right) + \Im\left( F_{T_1} F_S^{*}
\right) \Big) \bigg),\label{eq:T1}\\ %
T_2 &= -8 b^2 \left( \modulus{F_{A}^+}^2 + \modulus{F_{V}^+}^2 - s \left(
\modulus{F_{T_1}}^2+\modulus{F_{T_2}}^2 \right) \right),
\end{align}
\end{subequations}
where $a$ and $b$ are as given in Eq.~\eqref{eq:ab}. In the limit of $m_{\ell}
\to 0$ (which is a reasonable approximation to make at the $B$ meson mass
scale), we get
\begin{subequations}
\begin{align}
T_0 \Big|_{m_{\ell}=0} &= 8~b^2 \left( \modulus{F_{A}^+}^2 + \modulus{F_{V}^+}^2
\right) + 2~s \left( \modulus{F_P}^2 + \modulus{F_S}^2 \right),\\ %
T_1 \Big|_{m_{\ell}=0} &= 8~b~s~\Big( \Im\left( F_{T_2} F_P^{*} \right) +
\Im\left( F_{T_1} F_S^{*} \right) \Big),\label{eq:T1-ml-zero}\\ %
T_2 \Big|_{m_{\ell}=0} &= -8 b^2 \left( \modulus{F_{A}^+}^2 +
\modulus{F_{V}^+}^2 - s \left( \modulus{F_{T_1}}^2+\modulus{F_{T_2}}^2 \right)
\right),
\end{align}
\end{subequations}
where we have used the fact that $b^2 = a^2 - m_D^2 \left( 2~a + s - m_D^2
\right)$. It is easy to notice that in the approximation of massless electrons
and muons, it is the $T_1$ term which carries the interference terms. We shall
define three asymmetries $A_j$ ($j=0,1,2$) which would be proportional to the
terms $T_j$ in Eq.~\eqref{eq:diff-decay-rate}.
\begin{subequations}\label{eq:asymmetries}
\begin{align}
A_0 &= -\frac{1}{6} \left( \int_{-1}^{-1/2} - 7 \int_{-1/2}^{1/2} +
\int_{1/2}^{1} \right) \frac{d^2\Gamma}{ds \, d\cos\theta} \; d\cos\theta
\nonumber\\%
& = \frac{b~\sqrt{s}}{128~\pi^3~m_B^2 \left( m_B^2 - m_D^2 + s \right)}T_0~,\\ %
A_1 &= - \left( \int_{-1}^{0} - \int_{0}^{1} \right) \frac{d^2\Gamma}{ds \,
d\cos\theta} \; d\cos\theta \nonumber\\%
& = \frac{b~\sqrt{s}}{128~\pi^3~m_B^2 \left( m_B^2 - m_D^2 + s \right)}T_1~,\\ %
A_2 &= 2 \left( \int_{-1}^{-1/2} - \int_{-1/2}^{1/2} + \int_{1/2}^{1} \right)
\frac{d^2\Gamma}{ds \, d\cos\theta} \; d\cos\theta \nonumber\\%
& = \frac{b~\sqrt{s}}{128~\pi^3~m_B^2 \left( m_B^2 - m_D^2 + s \right)}T_2~.
\end{align}
\end{subequations}
It is important to notice that these asymmetries are defined in the
Gottfried-Jackson frame (in which the two leptons fly away from each other
back-to-back) and not in the laboratory frame. It is also easy to notice that
the forward-backward asymmetry $A_{FB}$ (again in the Gottfried-Jackson frame)
is related to the asymmetry $A_1$ as follows,
\begin{equation}
A_{FB} \equiv  \left( \int_{-1}^{0} - \int_{0}^{1} \right) \frac{d^2\Gamma}{ds
\, d\cos\theta} \; d\cos\theta= -A_1.
\end{equation}
Let us now analyse the three asymmetries keeping an eye on signatures of any new
physics.

\subsection{The model-independent signatures of new physics}\label{sec:new_physics_signatures}

In case of the SM, only $F_A$ and $F_V$ contribute to the angular distribution.
Therefore, all the interference terms in the expression for $T_1$ in
Eq.~\eqref{eq:T1} are identically equal to zero, making $T_1$ to vanish in the
SM. Moreover, if we consider the leptons to be massless, the combination
$\Big(T_0 + T_2 \Big)\Big|_{m_{\ell}=0} = 8b^2 s \left( \modulus{F_{T_1}}^2 +
\modulus{F_{T_2}}^2 \right) + 2s \left( \modulus{F_P}^2 + \modulus{F_S}^2
\right)$ also vanishes in the SM. Therefore, in the SM we have the following
predictions,
\begin{subequations}
\begin{align}
T_1 &= 0,&& \textrm{(SM prediction)}\\%
T_0 + T_2 &= 0. && \textrm{(SM prediction with $m_{\ell}=0$)}
\end{align}
\end{subequations}
Since, $A_n \propto T_n$ (for $n=0,1,2$) with the same constant of
proportionality as would be clear from Eq.~\eqref{eq:asymmetries}, we get the
following predictions for the asymmetries from the SM,
\begin{subequations}
\begin{align}
A_1 = -A_{FB} &= 0,&& \textrm{(SM prediction)}\\%
A_0 + A_2 &= 0. && \textrm{(SM prediction wth $m_{\ell}=0$)}
\end{align}
\end{subequations}
So the SM predicts that the forward-backward asymmetry in the decay modes under
our consideration be identically equal to zero. Moreover, for the SM, the $T_2$
term is given by
\begin{equation*}
T_2 = -8 b^2 \left( \modulus{F_{A}^+}^2 + \modulus{F_{V}^+}^2 \right),
\end{equation*}
and considering massless leptons the $T_0$ term is given by (in the SM),
\begin{equation*}
T_0 = 8 b^2 \left( \modulus{F_{A}^+}^2 + \modulus{F_{V}^+}^2 \right).
\end{equation*}
In presence of sizeable new physics contribution we find that
\begin{align}
& T_0 > 8 b^2 \left( \modulus{F_{A}^+}^2 + \modulus{F_{V}^+}^2 \right),\quad T_2
> -8 b^2 \left( \modulus{F_{A}^+}^2 + \modulus{F_{V}^+}^2 \right), \nonumber\\ %
\implies & T_0 + T_2 > 0.
\end{align}
Also in the presence of new physics
\begin{equation}
T_1 \neq 0.
\end{equation}
Translating these into the observables $A_0$, $A_1$ and $A_2$ we find the
following signatures of new physics (NP),
\begin{subequations}\label{eq:new-physics-signatures}
\begin{align}
A_1 = - A_{FB} &\neq 0, && \textrm{(NP signature)} \label{eq:NP-A1}\\%
A_0 + A_2 &\neq 0. && \textrm{(NP signature with $m_{\ell}=0$)} \label{eq:NP-A0pA2}
\end{align}
\end{subequations}
It is important to note that these signatures are truly independent of any
specific model for new physics. Moreover, if \eqref{eq:NP-A1} gets satisfied it
automatically ensures that \eqref{eq:NP-A0pA2} is also true, but not vice versa.
This is because of the fact that the $T_1$ term has interference terms in it,
see Eqs.~\eqref{eq:T1} and \eqref{eq:T1-ml-zero}. We would like to emphasize
that Eq.~\eqref{eq:NP-A1} is true irrespective of the mass of the lepton.
Moreover, at the $B$ meson mass scale both electron and muon are effectively
massless, and hence effectively $A_0 + A_2 \neq 0$ is a signatre of new physics.

Since in the SM, $T_1 = 0$, the angular distribution in the SM is completely
symmetric under $\cos\theta \leftrightarrow -\cos\theta$. Any asymmetry in the
angular distribution under $\cos\theta \leftrightarrow -\cos\theta$ exchange is,
therefore, a distinct signature of new physics.

\subsection{Experimental signatures of new physics}\label{sec:expt-sign}

\begin{figure}[hbtp]
\centering %
\includegraphics[width=0.95\linewidth, keepaspectratio]{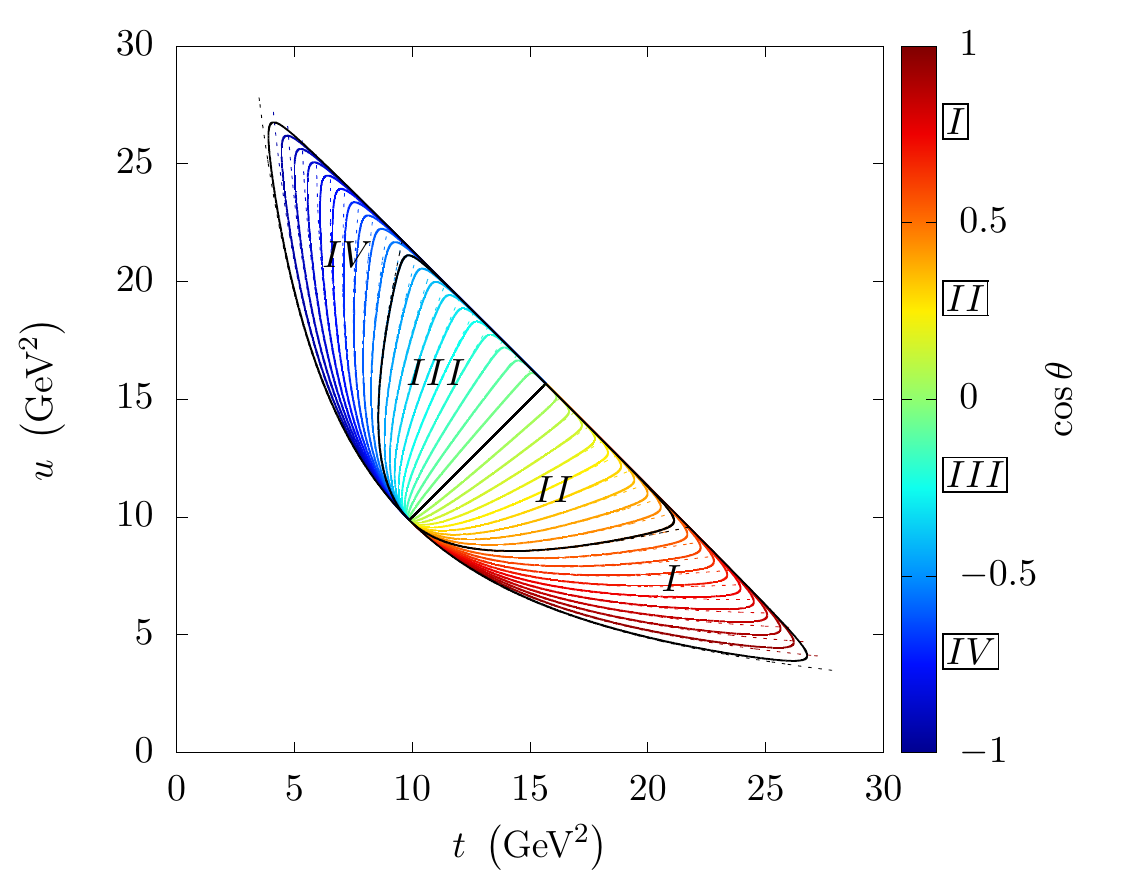} %
\caption{The region for Dalitz plot of $B \to D\mu^+ \mu^-$ showing the
variation of $\cos\theta$ inside it. The solid lines take into account the mass
of muon, but the dashed lines are for the massless muon case. The Dalitz plot
can be divided into four segments, denoted by $I$, $II$, $III$, and $IV$,
according to the region of $\cos\theta$ as shown along the color bar here. The
Dalitz plot can be obtained in any frame of reference. We need not go to the
Gottfried-Jackson frame for the Dalitz plot, the laboratory frame is
sufficient.} %
\label{fig:Dalitz-B2DMuMu} %
\end{figure}

Here we provide expressions for the asymmetries $A_0$, $A_1$ and $A_2$, in terms
of the easily observable distribution of events in the Dalitz plots for $B \to D
\ell^+ \ell^-$. The Dalitz plot can be obtained in any frame of reference, such
as laboratory frame. We note that the quantity $d^2 \Gamma/(ds~d\cos\theta)$
denotes the distribution of events inside the Dalitz plot. The
Fig.~\ref{fig:Dalitz-B2DMuMu} shows the variation of $\cos\theta$ inside the
Dalitz plot region allowed for $B \to D \mu^+ \mu^-$ decays. %
As shown in Fig.~\ref{fig:Dalitz-B2DMuMu} we can divide the Dalitz plot into
four regions as defined below,
\begin{align*}
&\textrm{Region }I: && 1 \geq \cos\theta \geq 0.5,\\%
&\textrm{Region }II: && 0.5 > \cos\theta \geq 0,\\%
&\textrm{Region }III: && 0 > \cos\theta \geq -0.5,\\%
&\textrm{Region }IV: && -0.5 > \cos\theta \geq -1.
\end{align*}
The asymmetries $A_0$, $A_1$ and $A_2$ can now be redefined as follows,
\begin{subequations}\label{eq:Dalitz-asymmetries}
\begin{align}
A_0 &= -\left( N_{I} - 7 N_{II} - 7 N_{III} + N_{IV} \right)/6,\\ %
A_1 &= N_{I} + N_{II} - \left(N_{III} + N_{IV}\right),\\ %
A_2 &= 2 \left( N_{I} - N_{II} - N_{III} + N_{IV} \right),
\end{align}
\end{subequations}
where $N_i$ denotes the number of events inside the Dalitz plot region $i$.
Since, the Dalitz plot for $B \to D \ell^+ \ell^-$ can be constructed in any
frame of reference, including the laboratory frame, it is easy to measure the
asymmetries as defined in Eq.~\eqref{eq:Dalitz-asymmetries}. The
model-independent signatures of new physics in terms of these three asymmetries
are given in Eq.~\eqref{eq:NP-A0pA2}.

We can also define another asymmetry which will probe the symmetry of
distribution of events in the Dalitz plot under $\cos\theta \leftrightarrow
-\cos\theta$ exchange. For this we need to divide the Dalitz plot into even
number of segments, each segment centered about some fixed value of
$\cos\theta$, say $c\theta_m$ and with width $\Delta c\theta$. The new
asymmetry, called the \textit{binned asymmetry} and denoted by $A_{\text{bin}}$
can be defined as follows,
\begin{equation}
A_{\text{bin}} = \sum_{c\theta_m} \frac{N(c\theta_m) -
N(-c\theta_m)}{N(c\theta_m) + N(-c\theta_m)},
\end{equation}
where $N(c\theta_m)$ denotes the number of events in the segment in which
$\cos\theta = c\theta_m \pm \Delta c\theta$. This binned asymmetry can be useful
in probing the symmetry of distribution of events in the Dalitz plot under
$\cos\theta \leftrightarrow -\cos\theta$ exchange. If experimentally
$A_{\text{bin}} \neq 0$, it would imply the presence of some NP. This asymmetry
would be more useful with large number of events in the Dalitz plot.

\subsection{Discussions regarding the effect of \texorpdfstring{$\Bz$-$\Bzb$}{B0-B0b} and \texorpdfstring{$\Dz$-$\Dzb$}{D0-D0b} mixing}\label{sec:mixing-independence}

It is essential to note that we get the same signatures of new physics as given
in Eq.~\eqref{eq:new-physics-signatures} whether we consider $\Bz \to \Dzb
\ell^+ \ell^-$ or $\Bz \to \Dz \ell^+ \ell^-$, thus implying that the $\DzDzb$
mixing has no effect on our analysis. Similarly, it is also true that
considering $\Bz \to \Dzb \ell^+ \ell^-$ or $\Bzb \to \Dzb \ell^+ \ell^-$ also
leads to the same signatures of new physics as given in
Eq.~\eqref{eq:new-physics-signatures}. Thus $\BzBzb$ mixing also plays no role
in our analysis. It must be noted that for each distinct decay mode under
consideration, the concerned quark currents are also distinct. However, for the
different $B \to D \ell^+ \ell^-$ decays (with their distinct quark currents),
we always get the same set of signatures for new physics as given in
Eq.~\eqref{eq:new-physics-signatures}. Thus for a quick search for signature of
new physics in the $B \to D \ell^+ \ell^-$ modes we can take any neutral $B$
meson as parent particle and consider any neutral $D$ meson in the daughter
particles. Furthermore, it must also be emphasized that even the events for $\Bp
\to \Dp \ell^+ \ell^-$ and $\Bm \to \Dm \ell^+ \ell^-$ decays can be added as
the charges of the $B$ and $D$ mesons do not affect the signatures of new
physics as given in Eq.~\eqref{eq:new-physics-signatures}. If we combine all
these decay modes, the data set will become larger and it will be possible to
get an early measurement of the signature of new physics. However, a mode
specific analysis would yield the extent to which new physics affects each
specific decay mode.

\section{Conclusions}\label{sec:conclusion}

Thus we have provided a fully model independent analysis of the rare $B \to D
\ell^+ \ell^-$ decays. We have provided the full model-independent expression
for the angular distribution along with expressions for three asymmetries which
are sensitive to the three distinguishable parts of the angular distribution. We
show that the three asymmetries are very sensitive to the presence of any new
physics. We have also provided the distinct signatures of new physics in terms
of these three experimentally observable asymmetries. Furthermore all the decay
modes can be analysed combinedly in the search for new physics, as $\BzBzb$ and
$\DzDzb$ mixings do not affect the concerned signatures of new physics as
enunciated in this paper. These features make this particular decay mode a very
interesting mode to look for in the various ongoing and upcoming $B$ physics
experiments, such as LHCb and Belle II.

\acknowledgments
The work of CSK was supported in part by the NRF grant funded by the Korea
government of the MEST \\(No.~2016R1D1A1A02936965). DS would like to thank
Prof.~Rahul Sinha for some helpful discussions regarding the decay mode at an
earlier stage of this work, and The Institute of Mathematical Sciences, Chennai,
India, where a part of this work was done, for hospitality.

\end{document}